\documentclass[12pt]{article}
\usepackage[utf8]{inputenc}
\usepackage{amsmath}
\usepackage{lineno,hyperref}
\usepackage{epsfig}
\begin{document}
\title{Procedure of simulations for Evaluations of Transient Parameters Stability for RF Cavities Due to Effects of Pulsed RF Heating}
\author{V.~Paramonov$^1$, B.~Militsyn$^2$, A.~Skassyrskaya$^1$\\
$ ^1$ - INR~of~the~RAS,~60-th~Oct.~Anniv.~Prosp.~7a,~117312,~Moscow,~Russia\\
$^2$ - UKRI~STFC~ASTeC,~Daresbury,~Warrington,~Cheshire,~UK }
\maketitle
\begin{abstract} 
The Pulsed RF Heating (PRFH) is the well known effect in the development of high gradient accelerating cavities. There 
is a lot of research dedicated to the tolerable strength of electromagnetic field and the corresponding temperature 
rise leading to the surface degradation and cavity performance violation. To provide high quality of electron 
bunches, RF gun cavities operate in electron sources with high electric, and hence, magnetic RF fields. Being at the 
safe side with respect to cavity surface degradation, related to the PRFH result for S band in essentially transient 
thermal deformations, leading to the cavity frequency shift and quality factor change even within a few 
microseconds in the RF pulse.\\
Thermo-elastic deformations consist of slow heat propagation from the cavity surface and fast elastic wave 
propagation from the very thin heated surface layer inside the cavity body with different in five orders typical 
velocities.\\
For simulations of PRFH effects for more or less practical cavity design we can get results only in 
direct numerical simulations with certified software like ANSYS. 
The critical issue for such direct simulations is to combine in a single model the possibility of correct description 
for such qualitatively different process and to prove the reasonable precision of simulations. 
This report describes the procedure for simulation of PRFH related transient thermo-elastic deformations in S band cavities.
The results of test simulations are presented to show the relative precision of simulations. 
\end{abstract}
\newpage
\tableofcontents
\newpage
\section{Introduction} 
\begin{figure}[htb]
\centering
\epsfig{file=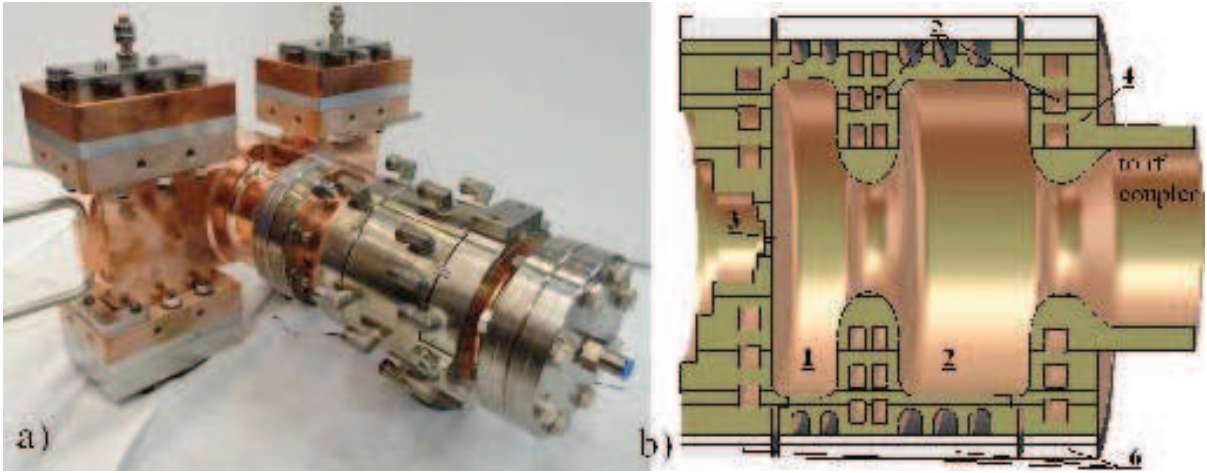, width =160.0mm}
\caption{The S band HRRG RF cavity for the CLARA project, (a) and the simplified model of the gun cavity. 1 - half cathode cell, 2 - full cell, 
3 - position of photo cathode, 4- cavity copper body, 5 - cooling channels,  6 - stainless steel jacket.}
\label{gun}
\end{figure}

The study of PRFH related effects usually is directed to estimations of the tolerable temperature rise at the 
cavity surface,~\cite{slac-pr}. To provide high quality of electron bunches for the modern Free Electron Laser (FEL) 
facilities, the RF gun cavities operate in the electron sources with high electric, and hence, magnetic RF fields. 
As a characteristic modern example Fig.~\ref{gun}a illustrates a High Repetition Rate Gun (HRRG) RF cavity 
of the CLARA project,~\cite{stfc}. The cavity operates with frequency $f=2998.5~MHZ$ in a pulsed mode with electric field at the cathode up 
to $E_c \sim 120~\frac{MV}{m}$. Certaimly this results in the maximal magnetic field at the surface 
$H_m \sim 250~\frac{kA}{m}$ and the maximal pulsed RF loss density up to $P_s \sim 4.5\cdot 10^8  \frac{W}{m^2}$. 
The PRFH effect during RF pulse $\tau \sim (3 \div 6)~ \mu s$ results in the temperature rise in some parts of the 
cavity surface at $T_s \sim~(19 \div 30)~C^o$. It is within the safe limits with respect to surface 
deterioration,~\cite{slac-pr}. But the temperature rise in the thin heated surface layer generates thermal 
deformations and change of cavity parameters during RF pulse. It will directly lead to deviations in the 
amplitude and the phase of RF field in the gun cavity.\\ 
The requirements to high performances for modern FEL facilities,~\cite{clara_des}, transform to the requirements of high 
stability of the field amplitude and phase in the RF gun cavity~\cite{jitter}. Estimations for parameters 
stability for each FEL element, including the RF gun, are required. The simplest and probably the only way to get 
estimations is the direct numerical simulations. For these reasons we need a reliable procedure and should represent
the precision of the results.  
\section{Process description}
The transient process of thermal elastic deformations is described by the coupled equations for temperature 
$T(\vec r,t)$ and displacements $\vec u(\vec r,t)$ distributions,~\cite{koval}:
\begin{eqnarray}
\nonumber
div(grad T(\vec r,t))-\frac{\rho_m C_p}{k_c} \frac{\partial T(\vec r,t)}{\partial t} 
- \frac{\alpha_t E_Y T_0}{3 k_c (1-2\nu)} \frac{\partial div \vec u(\vec r,t)}{\partial t}=0,\\
\frac{3(1-\nu)}{(1+\nu)}grad(div\vec u(\vec r,t)) -\frac{3(1-2\nu)}{2(1+\nu)}rotrot\vec u(\vec r,t) 
 -\frac{3(1-2\nu)\rho_m}{E_y} \frac{\partial^2 \vec u(\vec r,t)}{\partial t^2} = \alpha_t grad T(\vec r,t),
\label{1e}
\end{eqnarray} 
where $\rho_m, C_p, k_c, \alpha_t, E_Y$ and $\nu$ are the density, the specific heat, the heat conductivity, the coefficient 
of linear thermal expansion, the Young's modulus and the Poisson ratio, respectively.\\ 
The first equation in Eq.~\ref{1e}, neglecting the last term as small, describes the heat diffusion from the heated surface 
into the cavity body. This process is well studied and explained in~\cite{slac-pr} and related references. 
The temperature rise $T_s$ at the surface over the steady-state value and the heat penetration depth $D_d$ can be 
found,~\cite{wilson},~\cite{slac-pr}, as:
\begin{equation}
T_s=\frac{2P_{s} \sqrt{ \tau}}{\sqrt{\pi k_c \rho_m C_p}},\quad \alpha_d =\frac{k_c \tau}{\rho_m C_p}, \quad D_d=\sqrt{\alpha_d \tau},
\label{2e}
\end{equation} 
where $\alpha_d \approx 10^{-4}~\frac{m^2}{s}$ is the thermal diffusivity for copper. For the RF pulse 
$\tau =1~\mu s, ~3~\mu s,  ~6~\mu s$ and $ 10~\mu s$ the heat penetrates to the depth 
$D_d=10.7~\mu m,~18.5~\mu m,~26.2~\mu m$ and $~33.8~\mu m$, respectively. \\ 
The second equation in Eq.~\ref{1e} is a typical wave equation for displacements $\vec u(\vec r,t)$ with a driving 
term $\alpha_T grad T(\vec r,t)$. The equation describes propagation of longitudinal elastic waves of compression from 
the source - the heated layer at the surface - into cavity the body with the velocity in copper  
\begin{equation}
V_l=\sqrt{\frac{E_y (1-\nu)}{2 \rho_m (1+\nu)(1-2\nu)}} \approx 4700~\frac{m}{s}. 
\label{3e}
\end{equation} 
The typical distance from the heated cavity surface the cooling channels, providing violation of the cavity body 
homogeneity, is about  $\sim 3~mm$. It corresponds to the travel time for about $1.3~\mu s$ which is 
necessary for an elastic wave to go from the surface to the channel and back. With a typical duration of the RF pulse for S-band, frequency $f \sim 3000~MHz$, gun cavities 
$\tau \sim (3 \div 6)~\mu s$, during the main part of the RF pulse we should expect interaction of forward elastic 
waves, moving from the cavity surface into the body with backward waves, scattered at violations of homogeneity.\\
But the time of the RF pulse is not sufficient for formation of a steady-state distribution of deformations 
and the displacements $\vec u(\vec r,t)$ are essentially non-stationary.     
\section{Procedure of simulations}
The real cavity design, Fig.~\ref{gun}a, contains a lot of technical details, which are not determining for the 
problem. We have to consider a simpler model, taking into account exclusively essential components, 
Fig.~\ref{gun}b. Even for such a simplified model we can obtain only numerical results with direct numerical 
simulations by using the certified software like ANSYS,~\cite{ansys}.\\
Based on the previously developed procedure of coupled simulation~\cite{proc}, on the basis of ANSYS a special procedure was developed 
to simulate PRFH effects in the L-band cavities with frequency $f \sim 1300~MHz$. But L-band cavities operate with 
more than one order lower RF loss density $P_s$ and the PRFH effects take place mainly due to a long RF pulse 
$\tau \sim (800 \div 1000)~\mu s$. In this case the PRFH effects are relatively slow and the displacements can be considered 
in quasi-static approximation: 
\begin{eqnarray}
\nonumber
div(grad T(\vec r,t))-\frac{\rho_m C_p}{k_c} \frac{\partial T(\vec r,t)}{\partial t} =0,\\
\frac{3(1-\nu)}{(1+\nu)}grad(div\vec u(\vec r,t)) -\frac{3(1-2\nu)}{2(1+\nu)}rotrot\vec u(\vec r,t) 
 = \alpha_t grad T(\vec r,t).
\label{4e}
\end{eqnarray} 
It corresponds to the infinite velocity $V_l$ of elastic wave propagation. The comparison with the experiments for 
a PRFH induced 
cavity frequency shift shown reasonably good coincidence.\\
\begin{figure}[htb]
\centering
\epsfig{file=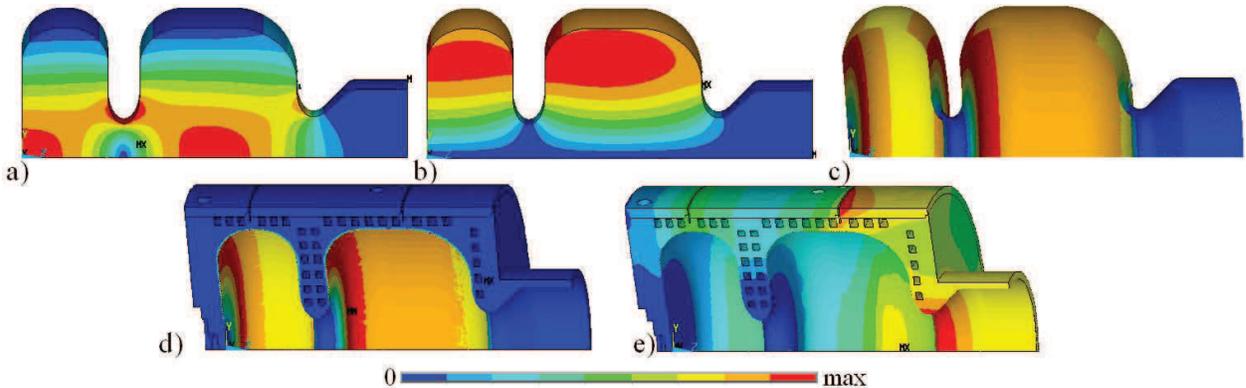, width =165.0mm}
\caption{Distributions of (a) electric, and (b) magnetic, fields in the gun cavity, (c) RF loss density at the 
surface, (d), 
temperature rise, and (e) displacements.}
\label{proc}
\end{figure}  
Fig.~\ref{proc} illustrates the procedure for PRFH effects simulations. Starting with simulation of electric and magnetic 
RF field distributions, Fig.~\ref{proc}a,b, we define the distribution of RF loss density at the cavity surface, Fig.~\ref{proc}c, 
and finally simulate temperature distribution in the heated layer, Fig.~\ref{proc}d and displacements distributions in the cavity 
body, Fig.~\ref{proc}e in quasi-static approximation, see Eq.~\ref{4e}.\\ 
Generally, the heat source is volumetric, distributed in a very thin layer with a skin-depth $\delta$ along the 
cavity surface. For L band frequencies $\delta \approx 1.4~\mu m$ and for S band $\delta \approx 0.9~\mu m$. 
As is shown in~\cite{slac-pr1}, in the equation for temperature distribution Eq.~\ref{1e} with a good precision of 
the results the boundary condition od the prescribed heat flux can be used:  
\begin{equation}
P_s(\vec r,t)=-\frac{1}{k_c} \frac{\partial T(\vec r,t)}{\partial n}.
\label{5e}
\end{equation} 
For a long RF pulse in L band cavities the approximation of the rectangular heat pulse can be used: 
\begin{equation}
P_s(\vec r,t)=0, t < 0, \quad t > \tau, \quad P_s(\vec r,t)=P_{s0}(\vec r), 0 \leq t \leq \tau,
\label{6e}
\end{equation} 
where $P_{s0}(\vec r)$ is the steady state distribution of RF losses at the cavity surface. For a relatively short 
RF pulse in the S band cavities we have to take into account the rise time of the field in the cavity $t_c$ and the 
pulse of heat load is accretive:
\begin{equation}
P_s(\vec r,t)=P_{s0}(\vec r)(1-e^{-\frac{t}{t_c}})^2=P_{s0}(\vec r)f_p(t) ,\quad 0 \leq t \leq \tau.
\label{7e}
\end{equation} 
For the HRRG gun cavity $t_c$ value is $ \approx 0.62~\mu s$.\\
In the quasi-static approximation for PRFH effects simulations for the L band cavities the first equation in 
Eq.~\ref{4e} is independent and the solution requires only one variable - temperature - per mesh node. The solution 
of the second equation in Eq.~\ref{4e} for displacements requires three independent variables per mesh node. 
For essentially transient deformations in the S band cavities we have to consider the coupled equations 
in Eq.~\ref{1e}, which require four variables per mesh node.\\
For the S band cavities PRFH effects take place due to an order higher heat loading $P_s$ and for the relatively short 
RF pulse the penetration depth $D_d$ is an order smaller. The distributions of $T(\vec r,t)$ and $grad T(\vec r,t)$ 
near the cavity surface are much sharper, as compared to the L band case. Hence, for the correct description of $T(\vec r,t)$ 
and especially $grad T(\vec r,t)$, much denser mesh is required near the cavity surface with not less than the second order for 
solution approximation.\\
To describe the body elements of the S band cavity even for a simplified model,  Fig.~\ref{gun}b, we need a mesh size 
$ h \leq 1.0~mm$ and for description of elastic wave propagation a time step $\delta t \sim \frac{h}{V_l} \sim 
0.2~\mu s$.\\
All these reasons together result in severe, about one order, increasing of the required resources in computing. 
Even for a simplified model in  Fig.~\ref{gun}b, being limited by problem size  $\sim 10^7$ variables, we cannot solve the task 
for the whole body and have to consider a sector of $30$ degrees.\\
For numerical simulations we have to define parameters of materials. For the cavity body, Fig.~\ref{gun}b, OFHC copper is applied.
Referring to~\cite{slac-pr}, the parameters of materials are assumed as is listed in Table~\ref{material_tab}.
\begin{table}[htb]   
\begin{center}
\caption{Material properties, used for simulations of coupled effects.}
\begin{tabular}{|l|c|c|c|c|c|c|c|c|c|c|c|c|c|c|c|c|}
\hline
Parameter                  & Units                   &  OFHC,                   & Steel              \\
                           &                         & annealed, \cite{slac-pr} & AISI 316           \\
\hline
Density, $\rho$            & $\frac{kg}{m^3}$        & 8950                     & 8000               \\ 
Specific heat, $C_p$       &$ \frac{J}{kg K^o}$      & 385                      & 460                \\
Heat conductivity, $k_c$   & $\frac{W}{m K^o}$       & 391                      & 16.3               \\
Thermal expansion, $\alpha$ & $\frac{1}{K^o}$         & $1.67 \cdot 10^{-5}$     & $1.59 \cdot 10^{-5}$  \\
Elastic modulus, $E_{Ym}$     & GPa                     & 123                      & 193                \\
Poisson's ratio, $\nu$      &                         & 0.345                    & 0.28                  \\
Yield stress, $\sigma_Y$   &  MPa                    & 62                       &                       \\
Electric conductivity, $\sigma$ & $\frac{S}{m}$      &   $5.8 \cdot 10^7$       &                       \\
\hline
\end{tabular}
\label{material_tab}
\end{center}
\end{table}     
\section{Test simulations}
\begin{figure}[htb]
\centering
\epsfig{file=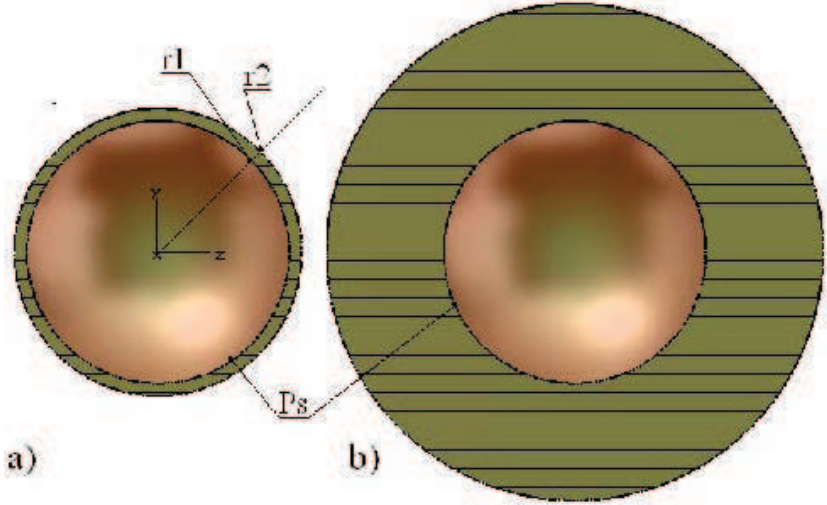, width =110.0mm}
\caption{Thin, a), and thick, b), hollow spheres for test simulations.} 
\label{spheres}
\end{figure}
\begin{figure}[htb]
\centering
\epsfig{file=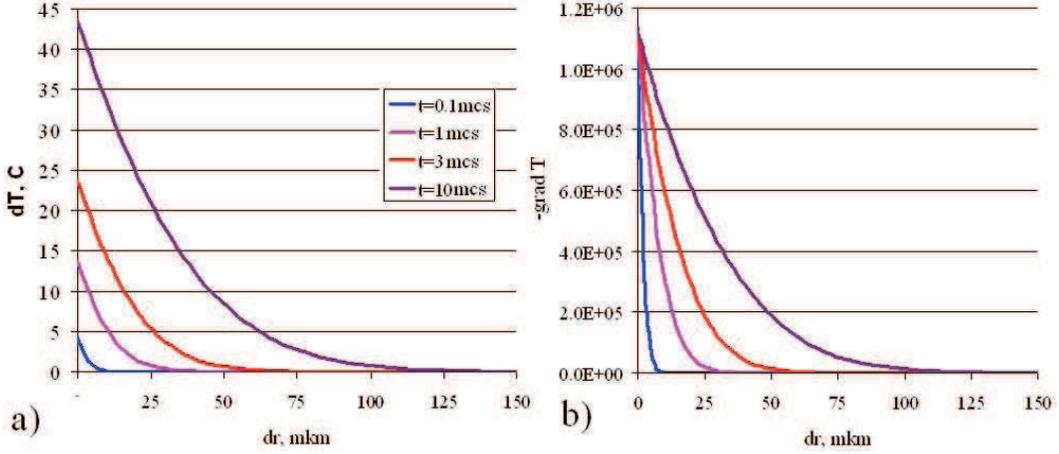, width =140.0mm}
\caption{Plots of temperature $T(r)$ distribution, a), and $grad T(r) =\frac{\partial T(r)}{\partial r}$, b) distribution for different time moments, $dr=r-r_1$ 
for the hollow sphere.}
\label{temp_grad_test}
\end{figure}

 
In any case, to be confident in the reliability of the results, for new physical conditions of PRFH, the procedure for 
simulations should be tested at some simpler model, where the results can be obtained with independent methods.\\
Let us consider hollow copper spheres as a test model, Fig.~\ref{spheres}, with the internal radius $r_1$ and outer radius $r_2$. 
For the thin sphere in Fig.~\ref{spheres}a $r_2-r_1=3~mm$ and for the time of simulations $t= 10~\mu s$ we can expect 
for an elastic wave several passes from the internal surface and back. For the thick sphere in Fig.~\ref{spheres}b 
$r_2-r_1=30~mm$
and for same time no reflected wave will occur at the inner surface.\\ 
We will also consider the uniform constant pulsed heat load $P_{s0} = 4.48\cdot 10^8  \frac{W}{m^2}$, applied to the inner sphere 
surface, $r=r_1$. 
In these approximations the coupled problem of thermal deformations, Eq.~\ref{1e}, in the spherical coordinates 
becomes one dimensional and solutions depend only on radius $r$: 
\begin{eqnarray}
\nonumber
\frac{1}{r^2} \frac{\partial}{\partial r}(r^2 \frac{\partial T(r,t)}{\partial r})-\frac{\rho_m C_p}{k_c} \frac{\partial T(r,t)}{\partial t} 
- \frac{\alpha_t E_Y T_0}{3 k_c (1-2\nu)}\frac{1}{r^2} \frac{\partial}{\partial t}(\frac{\partial (r^2 u_r(r,t))}{\partial r})=0,\\ 
\frac{3(1-\nu)}{(1+\nu)}\frac{1}{\partial r} (\frac{1}{r^2}\frac{\partial (r^2 u_r(r,t))}{\partial r})
 -\frac{3(1-2\nu)\rho_m}{E_y} \frac{\partial^2 u_r(r,t)}{\partial t^2} = \alpha_t \frac{\partial T(r,t)}{\partial r}.
\label{8e}
\end{eqnarray} 
The system of Eq.~\ref{8e} can be solved numerically by using routines from the program libraries like NAG,~\cite{nag}.  
Fig.~\ref{temp_grad_test}a,b shows the plots of temperature $T$ distribution and $grad T =\frac{\partial T}{\partial r}$ 
distribution inside the hollow sphere for different time moments. These results were obtained by numerical solution of Eq.~\ref{8e} 
using D03PBF routine in the NAG Fortran Library assuming the rectangular pulse of the heat load, Eq.~\ref{6e}. As one can see 
from Fig.~\ref{temp_grad_test}b, very sharp dependence for $grad T(r) =\frac{\partial T(r)}{\partial r}$ takes place in the pulse beginning,
$t \leq 1~\mu s$.  
The correct description of $grad T(r) =\frac{\partial T(r)}{\partial r}$ distribution due to PRFH effect is the main problem for numerical 
solution of Eq.~\ref{1e} problem in both transient and quasi-static approximations, because the expansion of the 
heated surface layer is the 
source of displacements. To estimate precision of $grad T(r)$ simulation, we can use the boundary condition in Eq.~\ref{6e}. 
Calculating $grad T(r)$ from the obtained $T(r)$ distribution for $r=r_1$ and comparing with the prescribed boundary condition 
$k_c P_s(t)=-\frac{\partial T(r,t)}{\partial r}$ 
we have estimations for precision of $grad T(r)$ calculation.\\  
For quasi-static approximation the second equation in Eq.~\ref{8e} has an analytical solution, regardless to details of 
temperature distribution $T(r)$. The equation for radial displacement $u_1$ in quasi-static approximation is:
\begin{equation}
\frac{1}{\partial r} (\frac{1}{r^2}\frac{\partial (r^2 u_r(r,t))}{\partial r}) =\frac{(1+\nu)\alpha_t}{3(1-\nu)} \frac{\partial T(r,t)}{\partial r}.
\label{9e}
\end{equation} 
It can be shown,~\cite{koval}, assuming $T(r)=0$ at the outer sphere surface $r=r_2$, that the azimuthal stress component 
$\sigma_{\varphi}(t)$ is:
\begin{equation}
\sigma_{\varphi}(t) = -\frac{E_Y \alpha_t}{(1-\nu)} \left ( T(r,t)-\frac{1}{r^3}\int_{r_1}^r T(r,t)r^2 dr +
\frac{1}{r_1^3-r_2^3}(2+\frac{r_1^3}{r^3})\int_{r_1}^{r_2} T(r,t)r^2 dr \right )=\sigma_{\vartheta}.
\label{10e}
\end{equation}
The inner, $r=r_1$, and outer, $r=r_2$ surfaces are free for expansion and the radial stress component $\sigma_r=0, r=r_1,r_2$. 
From the relation between the stress $\sigma$, deformations $\varepsilon$ and displacements $u$ in spherical 
coordinates, at the 
surfaces $r=r_1$ and $r=r_2$ we obtain: 
\begin{equation}
\frac{u_r(r,t)}{r}=\varepsilon_{\varphi}(r,t) = 
\frac{(1+\nu)\sigma_{\varphi}}{E_Y}- \nu \frac{\sigma_r+\sigma_{\varphi}+\sigma_{\vartheta}}{E_y} +\alpha_t T(r,t) =
\frac{(1-\nu)\sigma_{\varphi}}{E_Y} +\alpha_t T(r,t)
\label{11e}
\end{equation} 
From Eq.~\ref{11e} and Eq.~\ref{10e} for the displacement of the inner sphere surface $r=r_1$ we get:
\begin{equation}
u_r(r_1,t) =\frac{3 \alpha_t r_1}{r_2^3-r_1^3}\int_{r_1}^{r_2} T(r,t)r^2 dr,\quad u_r(r_2,t) =\frac{3 \alpha_t r_2}{r_2^3-r_1^3}\int_{r_1}^{r_2} T(r,t)r^2 dr.
\label{12e}
\end{equation}
But the integral in Eq.~\ref{12e} can be tranformed as:
\begin{eqnarray}
\nonumber
\int_{r_1}^{r_2} T(r,t)r^2 dr =  \frac{1}{4 \pi \rho_m C_p}\int_{r_1}^{r_2} \int_{0}^{\pi} \int_{0}^{2\pi} (\rho_m C_p T(r,t))r^2 cos{\vartheta}d\vartheta d\varphi dr=\\
 = \frac{1}{4 \pi \rho_m C_p} \int_{V} (\rho_m C_p T(r,t))dV = 
\frac{1}{4 \pi \rho_m C_p} \int_{S} \int_0^t P_s(r,t)dS dt = \frac{ r_1^2 P_{s0}}{\rho_m C_p} \int_0^t f_p(t) dt  
\label{13e}
\end{eqnarray} 
As is evident from Eq.~\ref{13e}, the integral $\int_{r_1}^{r_2} T(r,t)r^2 dr$ is proportional to the total heat amount, 
accepted by the body during the heat load pulse. The details of temperature distribution inside the body and the shape 
ofa  heat pulse are not essential.\\
The same conclusion for displacements of the inner cavity surface also was made in numerical simulations of PRFH 
effects for the L band cavities with much more complicated geometries,~\cite{l_pulse}.\\ 
With the transformation of Eq.~\ref{13e} the displacements at the inner sphere surface, see Eq.~\ref{12e}, become:
\begin{equation}
u_r(r_1,t) =\frac{3 \alpha_t r_1^3 P_{s0}}{\rho_m C_p (r_2^3-r_1^3)}\int_0^t f_p(t) dt, \quad u_r(r_2,t) 
=\frac{3 \alpha_t r_1^2 r_2 P_{s0}}{\rho_m C_p (r_2^3-r_1^3)}\int_0^t f_p(t) dt,
\label{14e}
\end{equation} 
and this relation can be used to compare the results and estimate precision of simulations with numerical methods.
\begin{figure}[htb]
\centering
\epsfig{file=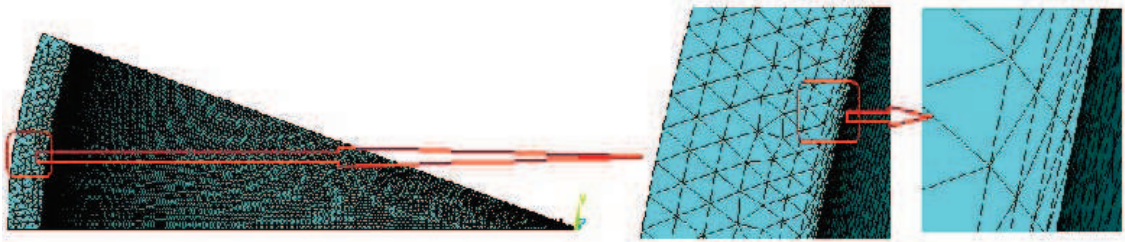, width =150.0mm}
\caption{FEM mesh for test simulations of PRFH effects in the hollow sphere.} 
\label{mesh_test}
\end{figure}
\subsection{Test simulations with ANSYS in quasi-static approximation}
\begin{figure}[htb]
\centering
\epsfig{file=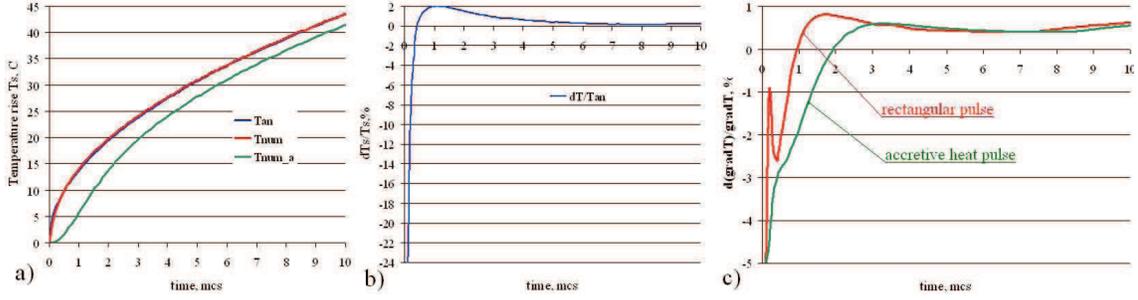, width =150.0mm}
\caption{Plots of the surface temperature rise $T_s$, (a), the relative error in $T_s$ for rectangular pulse, (b) and the relative errors 
in $grad T(r)$ simulations for rectangular and accretive heat pulses.}
\label{sphere_temp_grad}
\end{figure}

\begin{figure}[htb]
\centering
\epsfig{file=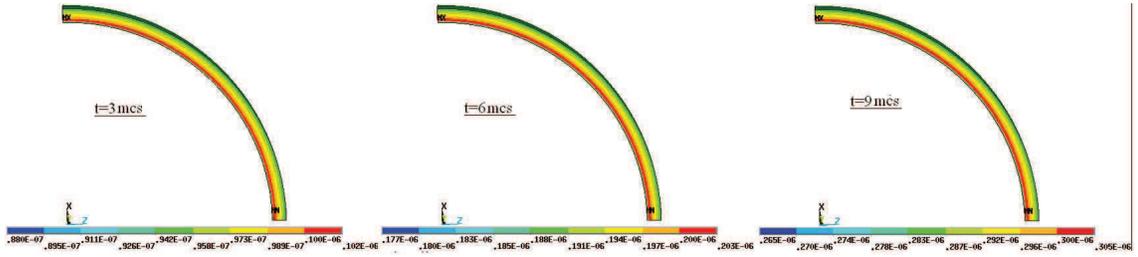, width =150.0mm}
\caption{Distributions of radial displacements inside the thin sphere for different time moments in quasi-static approximation 
and rectangular heat pulse.}
\label{sphere_stat_disp}
\end{figure}

\begin{figure}[htb]
\centering
\epsfig{file=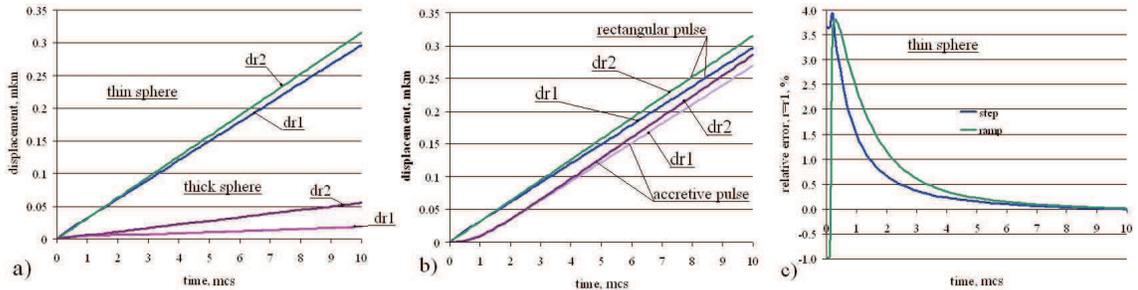, width =150.0mm}
\caption{Calculated with ANSYS displacements of inner, $dr_1$, and outer, $dr_2$, surfaces for thin and thick hollow spheres and 
rectangular heat pulse, (a), displacements for the thin sphere for rectangular and accretive heat pulses, (b) and relative error 
for $dr_1$ definition for rectangular (blue curve) and accretive (green curve) pulses, (c).}
\label{sphere_r1_r2_dr1}
\end{figure}

The ANSYS software is based on the Finite Elements Method (FEM) and the results of simulations, in general, depend on the mesh type and quality.
From the plots in Fig.~\ref{temp_grad_test}b one can see the necessity of $grad T(r) =\frac{\partial T(r)}{\partial r}$ approximations with 
several straight (linear) segments at a distance of about $\approx 2 D_d$. Hence, for approximation of temperature distribution a FEM with the second 
or higher order of approximation should be used. The mesh step in the direction normal to the surface should be as 
small as possible 
from free mesh generation realization. For the complicated real shape of the surface, see Fig.~\ref{gun}b, it is limited by precision of 
Boolean operations during numerical model design. The FEM mesh, generated for ANSYS test simulations of PRFH effects in the hollow sphere, 
is shown in Fig.~\ref{mesh_test}.\\   
Fig.~\ref{sphere_temp_grad}a shows the plots of the surface temperature rise $T_s$ obtained from 1D estimation 
Eq.~\ref{2e} for a rectangular heat pulse, Eq.~\ref{6e}, $T_{an}$ (blue curve) and obtained from ANSYS simulations for rectangular, $T_{num}$ (red 
curve) and accretive, $T_{numa}$, Eq.~\ref{7e} (green curve) pulses. The analytical 1D estimation, see Eq.~\ref{2e}, is precise for the 
flat surface but works very well if the local curvature radius $R_c \gg D_d$. In Fig.~\ref{sphere_temp_grad}b is shown The plot 
of relative error for temperature simulations $dT=\frac{T_{num}-T_{an}}{T_{an}}$ in Fig.~\ref{sphere_temp_grad}b is shown. Fig.~\ref{sphere_temp_grad}c 
depicts the plots of relative errors for $grad T(r,t)$ simulations, defined as 
$(\frac{\partial T_{num}(r,t)}{\partial r}-\frac{P_s(t)}{k_c})/(\frac{P_s(t)}{k_c})$
for rectangular (red curve) and accretive ( green curve) pulses. As we see from the plots in Fig.~\ref{sphere_temp_grad}b,c, in the pulse 
beginning $0 \leq t \leq (1 \div 1.5) \mu s$ there are some deviations between numerically calculated and precise values for both  
the temperature rise $T_s$ and for the first derivative - a temperature gradient $\frac{\partial T(r,t)}{\partial r}$. But these deviations 
are expressed in percent. For the rest of the pulse we see a very good coincidence in numerically calculated and precise values.\\
According to Eq.~\ref{12e} in quasi-static approximation the distribution of radial displacements in radius during the pulse changes 
only in the thin heat layer. In the rest of the body and at the surfaces, Eq.~\ref{14e}, the value of displacements for a rectangular 
pulse linearly rises with the time. It is well confirmed by calculated distributions for different time moments, presented in 
Fig.~\ref{sphere_stat_disp}.\\ 
Fig.~\ref{sphere_r1_r2_dr1}a depicts the plots of numerically calculated displacements of the inner $r=r_1$ and outer $r=r_2$ surfaces for 
thin and thick spheres and a rectangular heat pulse. One can see straight lines, which reflect linear proportionality to time or 
proportionality to the accepted amount of heat. Fig.~\ref{sphere_r1_r2_dr1}b illustrates the same displacements for inner and outer 
surfaces for the thin sphere, but shows comparison for rectangular and accretive heat pulses. In agreement with Eq.~\ref{14e} for an 
accretive pulse the displacements are delayed at a time $\approx 2 t_c$ as compared to a rectangular pulse. Relations Eq.~\ref{14e} can 
be used for comparison of numerically calculated and precise values of displacements at surfaces $r=r_1,r_2$ for both rectangular and 
accretive pulse shapes. 
Fig.~\ref{sphere_r1_r2_dr1}c presents the plots for relative deviations between the numerically calculated $dr_{1num}$ and precise, 
$dr_{1an}$, obtained from Eq.~\ref{14e}, values of the inner surface displacements $dr_1$ in the thin sphere. The deviations are defined 
as $\frac{dr_{1an}- dr_{1num}}{dr_{1an}}$. As we see from the plots in Fig.~\ref{sphere_r1_r2_dr1}c, only during the pulse beginning 
$0 \leq t \leq 3 t_c$ the deviations exceed $1\%$ for both rectangular and accretive pulse shapes. During this time period the values 
of displacements are less than $0.05~\mu m$ and there can be a significant influence of numerical errors in computing a big system with 
$\sim 10^7$ variables.  
\subsection{Test simulations with ANSYS in transient approximation}
\begin{figure}[htb]
\centering
\epsfig{file=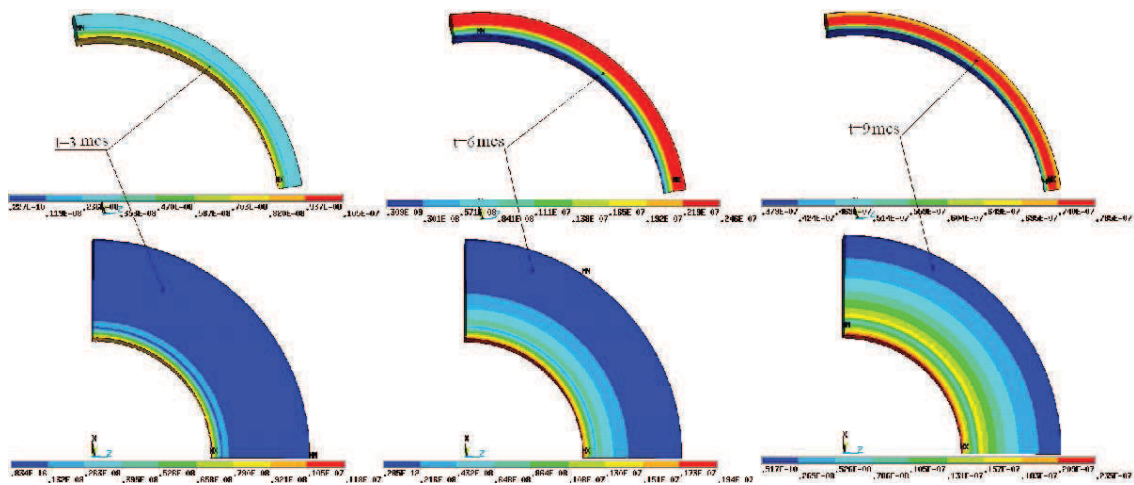, width =150.0mm}
\caption{Displacements distributions inside the thin (top row) and thick (bottom row) spheres for different time moments 
in transient approximation and a rectangular heat pulse.}
\label{usum_sphere_tran}
\end{figure}  
 The distribution of displacements in the transient approximation with propagating elastic waves strongly differs from 
the distribution in quasi-static approximation. Qualitatively the $u_r(r,t)$ displacements in transient approximations are 
shown in Fig.~\ref{usum_sphere_tran} for thin and thick spheres. For the thin sphere $r_2=r_1+3~mm$ the time of integration 
$\sim 10~\mu s$ is sufficient to see interference of a direct wave from the inner surface with a backward wave, reflected from the outer 
surface. In Fig.~\ref{usum_sphere_tran} ( top row) we can distinguish at the inner surface the dominating forward wave at 
$0 \leq t \leq 3~\mu s$ and interaction with the backward wave for $3 \leq t \leq 10~\mu s$. For the thick sphere $r_2=r_1+30~mm$ in 
Fig.~\ref{usum_sphere_tran} (bottom row) we see only propagation of the direct wave from the inner surface.\\ 
\begin{figure}[htb]
\centering
\epsfig{file=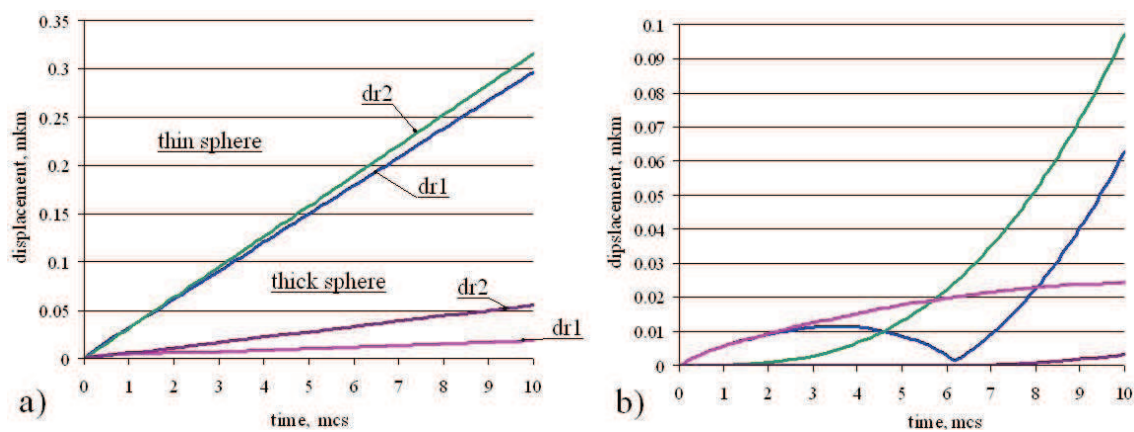, width =150.0mm}
\caption{Displacements of inner, $dr_1$, and outer, $dr_2$, surfaces for thin and thick spheres in static, (a) and transient, (b) 
approximations for rectangular heat pulse.}
\label{sphere_u_stat_tran}
\end{figure} 
To compare displacements in numbers, Fig.~\ref{sphere_u_stat_tran}a reproduces the same plots as in Fig.~\ref{sphere_r1_r2_dr1}a 
- the plots of surfaces displacements $dr_1$ and $dr_2$ in quasi-static approximation for both thin and thick spheres. The plots in the 
same colors for the corresponding displacements are shown in Fig.~\ref{sphere_u_stat_tran}b for transient approximation. In transient 
approximation we do not have any solution for displacements reference  like Eq.~\ref{14e} for the quasi-static case. But qualitatively 
transient displacements are logically consistent.\\   
For the thin sphere at the inner surface, Fig.~\ref{sphere_u_stat_tran}b (blue curve) we see oscillation of displacement $dr_1$, 
which strongly differs from the straight line in Fig.~\ref{sphere_u_stat_tran}a. For the thick sphere we see the growing with time 
and going to saturation displacement at the inner surface and the beginning of a displacement growth for the outer surface.\\
The test simulations for a simplified model and the comparison with precise solutions for temperature distribution $T(r,t)$, the description 
of the thin heated layer with $grad T(r_1,t)$ and finally the displacements of the surfaces have shown a very good coincidence during the 
main part 
of the heat pulse. In the pulse beginning there are deviations between numerically calculated and precise values, but in numbers these 
relative deviations are in percentage units. Such a precision is quite sufficient to consider the numerical results as reasonably reliable 
and precise estimations.    
\section{Application to PRFH effects simulations in the S band cavity}
The same strategy with the same requirements to mesh generation was applied to PRFH effects simulations with ANSYS in the S band cavity, 
Fig.~\ref{gun}b.
\begin{figure}[htb]
\centering
\epsfig{file=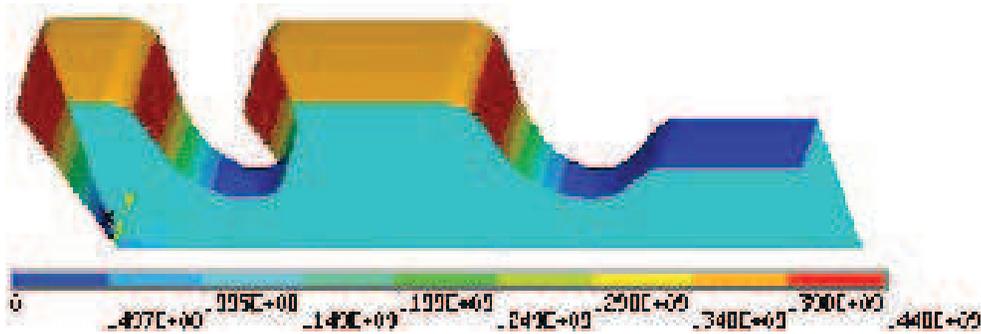, width =130.0mm}
\caption{Distribution of RF loss density at the cavity surface.}
\label{hflu}
\end{figure}  
The calculated distribution of RF loss density at the cavity inner surface is shown in Fig.~\ref{hflu}.\\ 
\begin{figure}[hb]
\centering
\epsfig{file=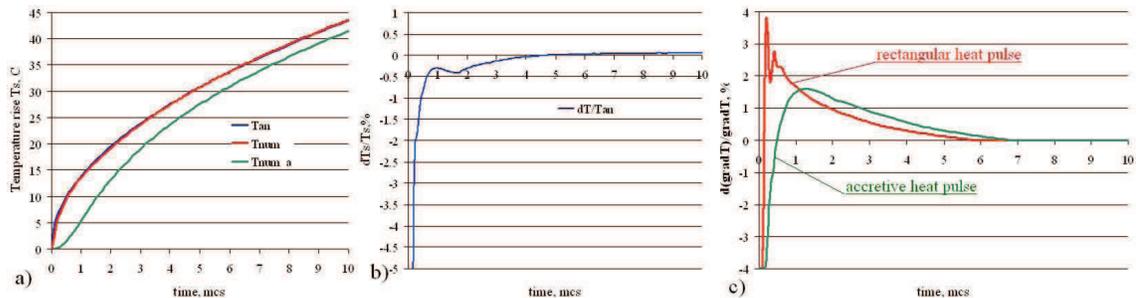, width =150.0mm}
\caption{Plots of the maximal surface temperature rise $T_{smax}$, (a), the relative error in $T_{smax}$ for a rectangular pulse, (b) and the 
relative errors in $grad T(r,t)$ simulations for rectangular and accretive heat pulses, (c).}
\label{gun_temp_grad}
\end{figure}  
The calculated distribution of a temperature rise $T_s$ at the inner cavity surface with the precision of $\sim D_d$ reflects the 
distribution of RF loss density.\\
For comparison of the maximal temperature rise $T_{smax}$ and $grad T$ at the cavity surface we 
can use both the 1D estimation, see Eq.~\ref{2e}, and the sequence of boundary conditions  $(\frac{\partial T_{num}(r,t)}{\partial r}-\frac{P_s(t)}{k_c})/(\frac{P_s(t)}{k_c})$.
Appropriate plots are shown in Fig.~\ref{gun_temp_grad}, similar to the corresponding plots in Fig.~\ref{sphere_temp_grad} from test 
simulations with spheres.\\    
Instead of more complicated geometry of the cavity surface, comparing the values for relative deviations in $T_s$  and $grad T$ numerical 
simulations we see practically the same numbers for both test spheres, Fig.~\ref{sphere_temp_grad}, and the realistic cavity 
surface, Fig.~\ref{gun_temp_grad}. \\
It means the similar, as for test simulations, precision in simulations of surface displacements for the real cavity shape.\\
\\ 
The physical consequences of surface deformations, estimated for typical geometry of the S band RF gun cavity with a typical operating regime. 
 due to PRFH effects, are considered preliminary in~\cite{rf_trans}.
\section{Summary}
To investigate the parameters deviations for a real cavity design with numerical calculations, we have to use modern software 
for multiphysics simulations. But we have to be confident in the reliability and accuracy of the results, which depend 
on the simulations procedure and the parameters of numerical models.\\
The focus of this report is in determination of the numerical models and illustration of the precision of the results 
in simulation of coupled, but very different in typical parameters physical processes. To show and estimate reliability, 
the numerical results of test simulations for selected simplified models are compared with analytical solutions.\\
As a result, the described procedure of numerical simulations for real S band cavities provides reliable results for 
estimations of the PRFH effects influence on cavities parameters.   
\section{Acknowledgement}
The work is supported by STFC contract N 4070193704.

\end{document}